\documentclass{aa}  

\usepackage{graphicx}
\usepackage{txfonts}
\usepackage{lipsum}
\usepackage{subcaption}         
\usepackage{lscape}             
\usepackage{placeins}           

\usepackage{graphicx}
\usepackage{txfonts}
\usepackage[colorlinks=true, citecolor=blue, urlcolor=blue]{hyperref}
\usepackage{placeins}
\usepackage{color, soul}
\usepackage{ulem}
\usepackage{natbib}
\usepackage{lipsum}
\bibpunct{(}{)}{;}{a}{}{,} 
\usepackage{tabularx}  
\usepackage{caption}

\newcommand{\fluxcgs}{\ensuremath{\mathrm{erg}\,\mathrm{s}^{-1}\,\mathrm{cm}^{-2}}}

\begin{document}

   \title{HESS J1832-085: evidence for a new gamma-ray binary candidate}


%
     \authorrunning{De Sarkar $\&$ Ghosh}
     \titlerunning{A new candidate gamma-ray binary HESS J1832-085}

   \author{Agnibha De Sarkar\inst{1}
          \and Tanuman Ghosh\inst{2}
          }

   \institute{Institute of Space Sciences (ICE, CSIC), Campus UAB, Carrer de Can Magrans s/n, 08193 Barcelona, Spain\\
   \email{desarkar@ice.csic.es}
   \and
   Inter-University Centre for Astronomy and Astrophysics, Post Bag 4, Ganeshkhind, Pune 411007, India}

   \date{Received XXXX; accepted YYYY}

   \date{Received XXXX; accepted YYYY}

 
  \abstract
   {The Galactic plane survey conducted by the High Energy Stereoscopic System (H.E.S.S.) has revealed numerous teraelectronvolt (TeV) sources, many of which remain unidentified. HESS J1832-085 is a point-like TeV source lacking a confirmed multiwavelength (MWL) counterpart. In this paper, we present {evidence supporting a gamma-ray binary interpretation of HESS J1832-085}.}
   {We aim to investigate the nature of HESS J1832-085 using the available X-ray and {\it Fermi} - Large Area Telescope (LAT) data, complemented by near-Infrared (NIR) identification of a possible stellar companion and broadband radiative modeling, to assess its classification as a potential gamma-ray binary.} 
   {Archival X-ray observations were examined to search for a counterpart and to characterize its spectrum and potential variability. We further analyzed $\sim$17.3 yrs of {\it Fermi}-LAT data between 1 and 300 GeV to establish a possible gigaelectronvolt (GeV) counterpart of HESS J1832-085. We also examined archival NIR images from the UKIRT Infrared Deep Sky Survey (UKIDSS) {and the Two Micron All Sky Survey (2MASS), together with 2MASS catalog photometry.} The broadband emission was interpreted using models commonly applied to gamma-ray binaries.}
   {A potential X-ray counterpart is identified in archival X-ray data, exhibiting a hard, absorbed spectrum and moderate variability. We also {identify a tentative} GeV gamma-ray excess spatially consistent with HESS J1832-085 and the {proposed} X-ray counterpart. We further identified a candidate stellar counterpart in {both} the UKIDSS {and 2MASS} images, located in close spatial proximity to the X-ray counterpart. {The adopted illustrative binary radiative model broadly reproduces the X-ray and TeV emission, but fails to explain the emission from the tentative GeV excess.}}
   {Our results {support HESS J1832-085 as a promising gamma-ray binary candidate}, motivating dedicated MWL follow-up observations to confirm the source nature.}

   \keywords{binaries: general - Gamma rays: general - X-rays: binaries - Stars: individual: HESS J1832-085
               }

   \maketitle
\nolinenumbers

\section{Introduction}


Gamma-ray binaries are rare high-energy (HE) systems \citep{dubus13} consisting of a compact object, typically a neutron star, orbiting an O- or Be-type companion. 
In pulsar-powered systems, the relativistic wind, produced by the dissipation of the rotational energy of the pulsar, collides with the stellar wind \citep{Maraschi_81, Dubus_06, Huber_21}, forming a wind collision region, which {separates} both winds \citep{Bogavalov_08, Bosch_12, Huber_21}, where particles can be efficiently accelerated, likely via diffusive shock acceleration \citep{Sironi_11, Huber_21}. 
The resulting relativistic particles produce non-thermal emission primarily through inverse-Compton (IC) and synchrotron processes \citep{dubus13}. 
Alternative emission scenarios may arise depending on the nature of the compact object, the companion star, and the system geometry.
Confirmed and candidate gamma-ray binaries are generally characterized by soft teraelectronvolt (TeV) spectra and hard, strongly absorbed X-ray emission.

HESS J1832-085 is a point-like ($\sim 0.02^{\circ} \pm 0.012^{\circ}$) tera-electronvolt (TeV) source discovered in the Galactic plane survey by the High Energy Stereoscopic System (H.E.S.S.) \citep{hess18} at Galactic longitude $l = 23.21^{\circ}$ and latitude $b = 0.29^{\circ}$. 
A radio pulsar PSR J1832-0827 \citep{clifton86} at a distance of $4.4-6.1$ kpc \citep{frail91, cordes02}, with a spin-down luminosity $\dot{E} = 9.3 \times 10^{33} \rm \ erg \ s^{-1}$ \citep{hobbs04}, braking index $n = 2.5 \pm 0.9$ \citep{johnston99} and a characteristic age $\tau_c \approx 200 \ \rm kyr$ was detected to be spatially coincident with the H.E.S.S. source.
Additionally, another millisecond pulsar, PSR J1832-0836 at a distance of 1.1 kpc, with a 2.7 ms period \citep{burgay13}, $\dot{E} = 1.7 \times 10^{34} \ \rm erg \ s^{-1}$, and $\tau_c = 5 \times 10^9$ yr was also detected in the region \citep{cordes02}.
Notably, a supernova remnant (SNR) G23.11+0.18, detected in radio, was found to be spatially consistent with the H.E.S.S. source, at a distance of $4.6 \pm 0.8$ kpc \citep{maxted19}.
AX J1832.3-0840, most likely an intermediate polar (IP) \citep{kaur10}, and a steep-spectrum radio point source \citep{maxted19}, both of which are unlikely to be associated with the H.E.S.S. source, are also present in the region.
Additionally, a highly absorbed and hard-spectrum X-ray source was found near the centroid of HESS J1832-085 from the analysis of {\it Suzaku} data \citep{aktekin25}.
The X-ray source was interpreted as a possible pulsar wind nebula; however, this interpretation is not without its shortcomings \citep{aktekin25}.

In this paper, we investigate the nature of HESS J1832-085.
Sect. \ref{results} presents the results of our analyses, with Subsect. \ref{XMM} focusing on the X-ray counterpart of HESS J1832-085 and Subsect. \ref{4FGL} discussing its possible gigaelectronvolt (GeV) counterpart.
These subsections also describe the corresponding X-ray and \textit{Fermi} - Large Area Telescope (LAT) data analyses.
{Subsect. \ref{sec:nir} presents the candidate near-Infrared (NIR) counterpart and the photometric consistency checks.}
In Sect. \ref{discussion}, we interpret the resulting broadband emission within models commonly considered for gamma-ray binaries.
Taken together, our results provide evidence in favor of HESS J1832-085 being a gamma-ray binary candidate.

\section{Results}\label{results}

\subsection{X-ray counterpart of HESS J1832-085}\label{XMM}

\begin{figure}
    \centering
    \includegraphics[width=\columnwidth]{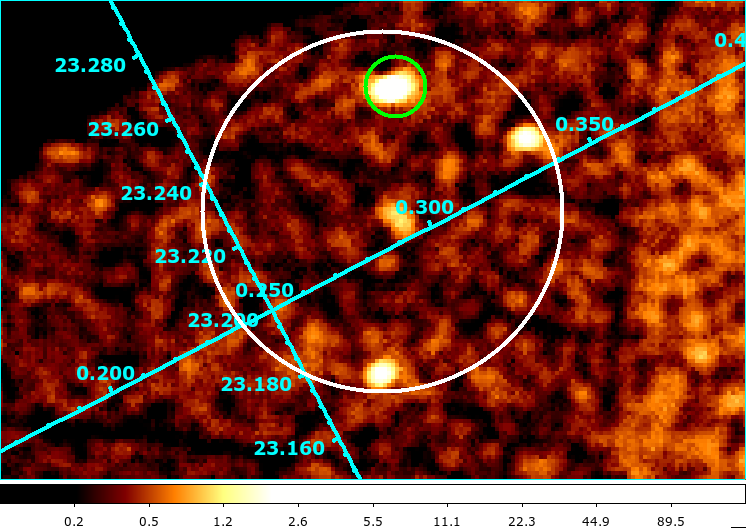}
    \caption{{{\it XMM-Newton} MOS2 image ($0.3\hbox{--}10.0$ keV) depicting the source region with a circle of $30^{\prime\prime}$ in green ($l = 23.2390621^{\circ}, b = 0.3090542^{\circ}$). This coordinate is close to the analysis of \cite{aktekin25} with much better constraint owing to a better spatial resolution of {\it XMM-Newton} detectors compared to that of {\it Suzaku}. The white circle, denoting the 95$\%$ positional uncertainty, is centered on HESS J1832-085 centroid \citep{hess18} and has a radius of $0.05^{\circ}$. The image is smoothed with a Gaussian kernel for visual purposes. The cyan grid shows the Galactic coordinates.}}
    \label{fig: XMM_image}
\end{figure}

\begin{figure}
    \centering
    \includegraphics[width=\columnwidth]{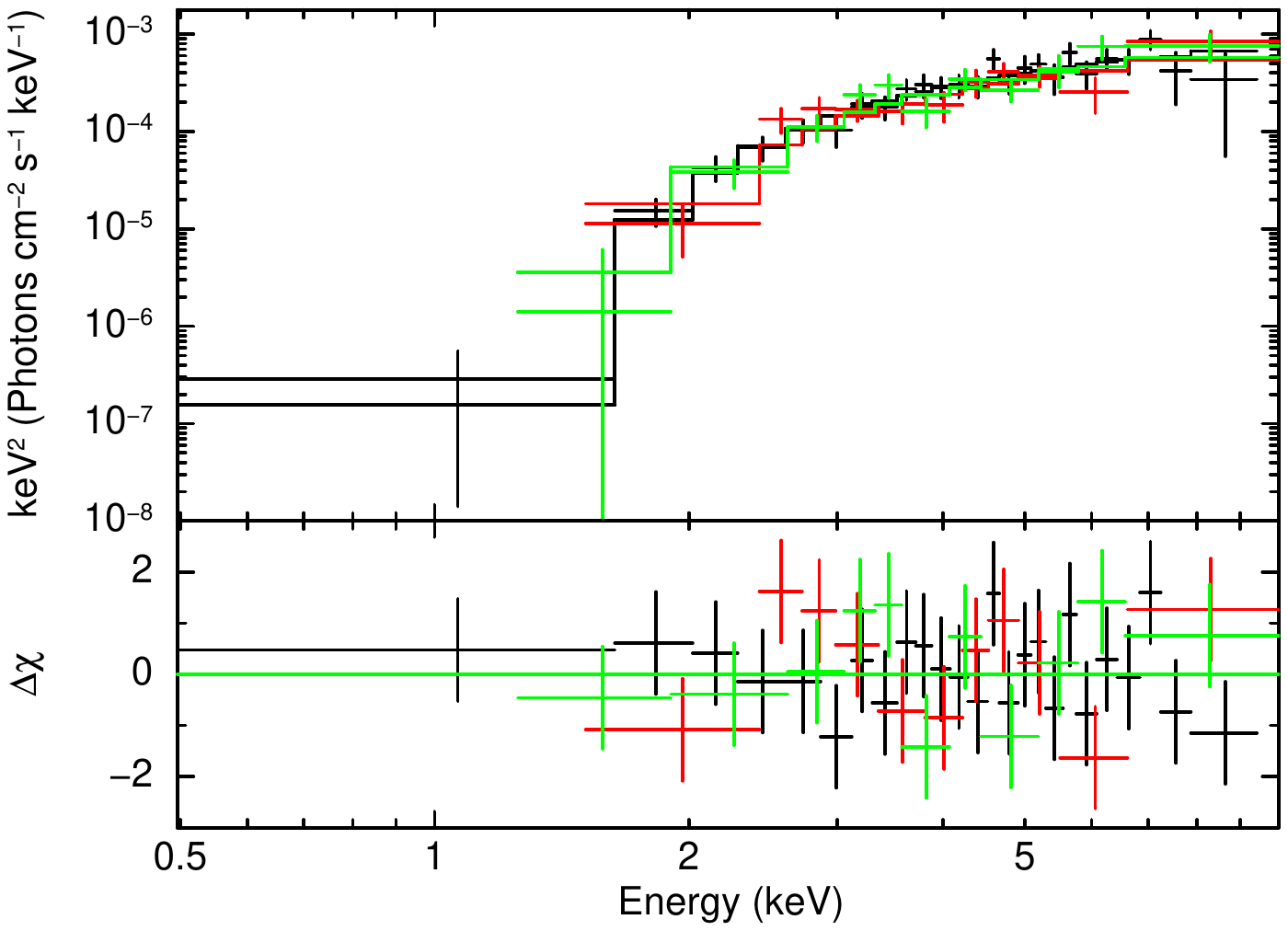}
    \caption{{The spectra and residuals (normalized by the statistical errors)} of the source with a simple absorbed power law fit to the {\it XMM-Newton} observation. pn, MOS1, and MOS2 are depicted with black, red, and green, respectively. Spectra are rebinned for visual purposes.}
    \label{fig: XMM_spectra}
\end{figure}

The X-ray counterpart of HESS J1832-085 was previously investigated using \textit{Suzaku} observations by \cite{aktekin25}.
Here, we extend this study using archival observations from \textit{XMM-Newton}, \textit{Chandra}, and \textit{Swift} to characterize the source spectrum and investigate its long-term X-ray variability.
The data extraction and analysis methodology is standard, using SAS v22.1.0 for {\it XMM-Newton}, HEASOFT v6.35.1 for {\it Swift}, and CIAO v4.15 for {\it Chandra}.
The \textit{XMM-Newton} data, which provide the best photon statistics among the available observations, are used for the detailed spectral analysis and subsequently for the multiwavelength (MWL) spectral energy distribution (SED) modeling, while the \textit{Chandra} and \textit{Swift} observations are primarily used for the study of long-term flux variation.
The primary data analysis methodologies are discussed below.

For the \textit{XMM-Newton} observation, the standard data analysis pipelines \texttt{epproc} and \texttt{emproc} for pn and MOS, respectively, are utilized, and the events are corrected for the background flaring using the \texttt{espfilt} task. The spectra are generated using the \texttt{evselect} tool with filters PATTERN<=4 for pn and PATTERN<=12 and FLAG==0 for all CCDs. The energy response matrix and ancillary files are generated using \texttt{rmfgen} and \texttt{arfgen} tasks. The spectra are grouped to a minimum $20$ counts per bin and an oversampling factor of $3$ using \texttt{specgroup}. The source and background regions are selected using $30^{\prime\prime}$ radius circles on the same chip.

For the \textit{Chandra} observation, the standard \texttt{chandra\_repro} task is utilized to obtain the reprocessed event file, and then spectra from source and background are extracted from $40^{\prime\prime}$ radius circles using \texttt{specextract}. Such a large extraction region is required because the source in that observation (ID - 11114) is very far off-axis, resulting in a broadened point spread function (PSF). The spectra are grouped until the number of counts in each group exceeds $1$, and while spectral fitting, \texttt{C statistic} has been used. 

For the \textit{Swift}-XRT observations, data were extracted using the standard \texttt{xrtpipeline} tool. After obtaining the event files, we further extracted the spectra from a $30^{\prime\prime}$ circle for the source, and a $60^{\prime\prime}$ circle for the background region. The spectra are grouped to $1$ count per bin, and while spectral fitting, \texttt{C statistic} has been used. We find that, for both {\it Swift} observations, the spectral parameters are consistent with those from the {\it XMM-Newton} data, but with significantly larger uncertainties due to low statistics. Therefore, we fix the absorption and power law index parameters to the best-fit value from {\it XMM-Newton}, to better constrain the flux.

Figure \ref{fig: XMM_image} shows the \textit{XMM-Newton} image of the region surrounding HESS J1832-085.
Within the 95\% positional uncertainty region of the H.E.S.S. source, the X-ray source marked by the green circle is the brightest detected source and coincides with the source previously proposed as a possible X-ray counterpart by \cite{aktekin25}.
Its close spatial proximity to the H.E.S.S. centroid therefore motivates a more detailed investigation of its properties to establish a plausible relation between this source and HESS J1832-085.

The analysis of {\it XMM-Newton} data shows that the source is spectrally hard and is well fitted with a simple absorbed power law model, as shown in Fig. \ref{fig: XMM_spectra}.
Using the \texttt{tbabs} model with the relevant elemental abundances \citep{Wilms} and absorption cross-section \citep{Verner}, we obtain a column density of $\rm N_H = 8.0^{+2.5}_{-2.0} \times 10^{22}$ cm$^{-2}$,
indicating that the neutral absorption parameter is high, and substantial absorption is expected along the line-of-sight.
The best-fit power law photon index is $\Gamma = 1.2^{+0.5}_{-0.4}$, demonstrating that the source has a hard X-ray spectrum, while the corresponding unabsorbed flux in the $2.0\hbox{-}10.0$ keV band is
$1.2^{+0.2}_{-0.1} \times 10^{-12}$ \fluxcgs.
The combination of a hard spectrum, strong absorption, and close positional agreement with HESS J1832-085 makes this source a plausible X-ray counterpart of the TeV source.

\begin{figure}
    \centering
    \includegraphics[width=\columnwidth]{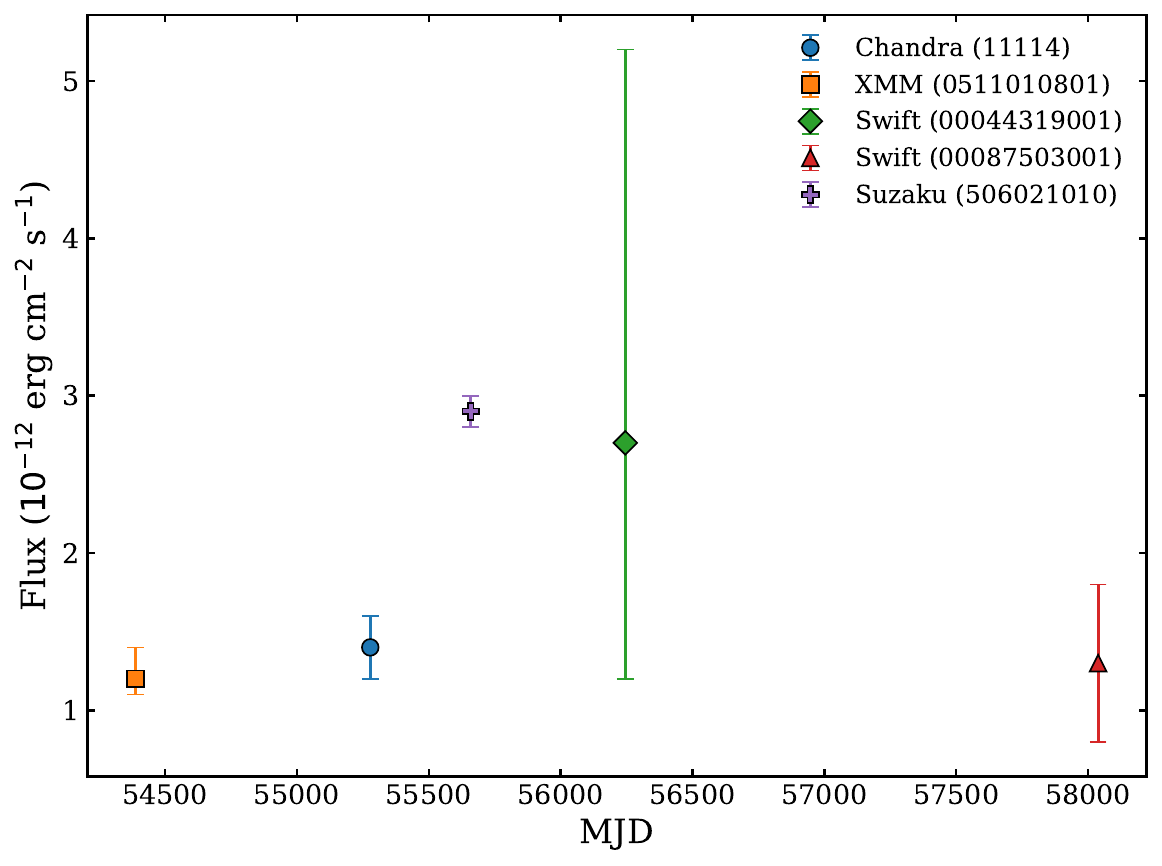}
    \caption{{Long-term unabsorbed $2.0 \hbox{-} 10.0$ keV flux variation over time from different observatories, depicting a flux variability in the source.}}
    \label{fig: xray_variab}
\end{figure}

We further investigated the long-term X-ray behavior of the source using the available \textit{XMM-Newton}, \textit{Chandra}, \textit{Swift}, and \textit{Suzaku} observations.
We have fitted the {\it XMM-Newton}, {\it Chandra}, and {\it Swift} spectra with a simple absorbed power law and estimated their fluxes in the $2.0\hbox{-}10.0$ keV energy range.
For \textit{Suzaku}, we adopted the flux reported by \citet{aktekin25}.
The resulting long-term unabsorbed $2.0\hbox{-}10.0$ keV flux variation over time is shown in Fig. \ref{fig: xray_variab}, with the corresponding observation IDs indicated for each measurement.
The measurements reveal moderate long-term flux variability, providing additional motivation for continued monitoring of the source and for investigating its possible association with HESS J1832-085.

\subsection{GeV counterpart of HESS J1832-085}\label{4FGL}

\begin{figure*}
     \includegraphics[width=\columnwidth]{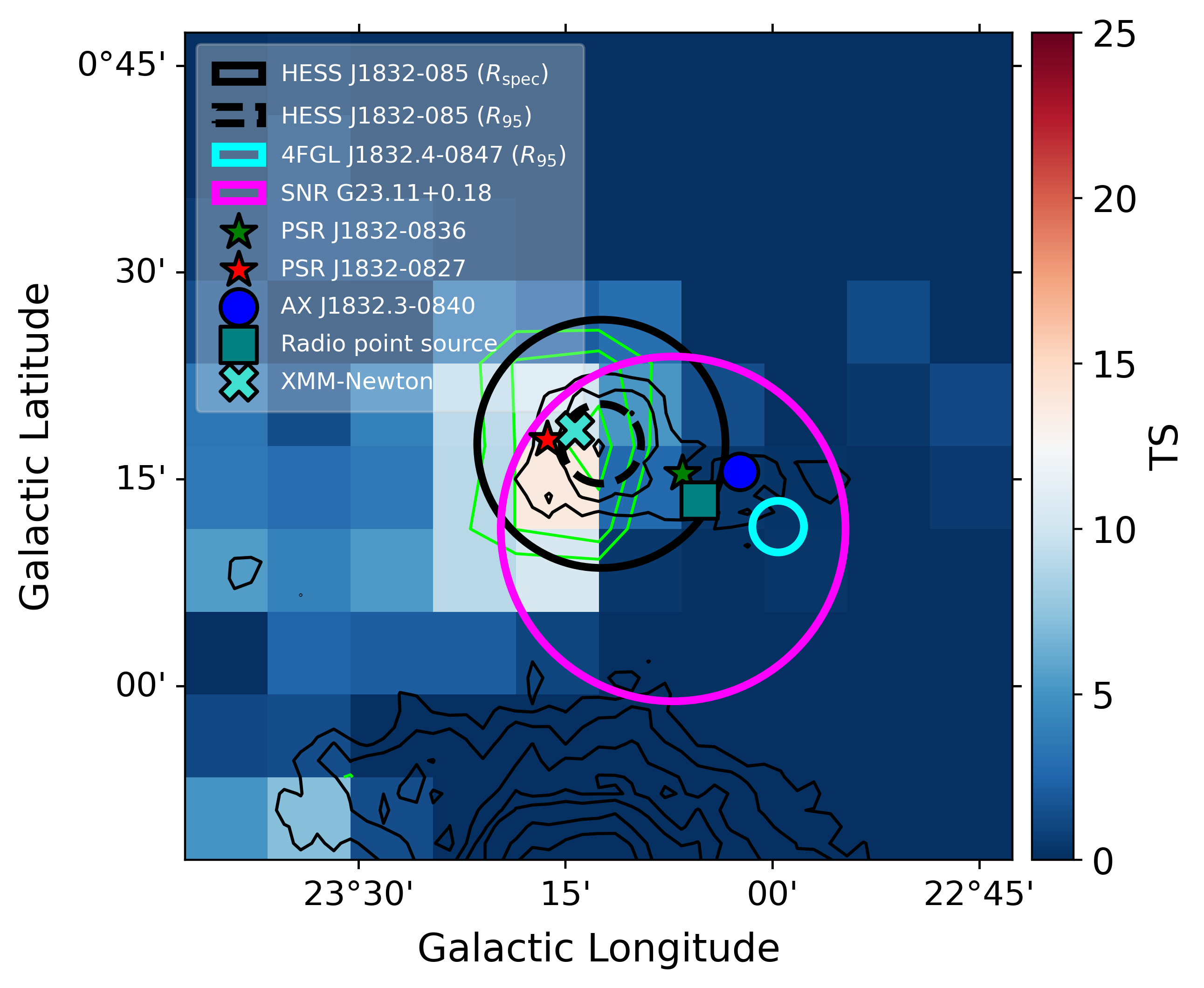}
     \includegraphics[width=\columnwidth]{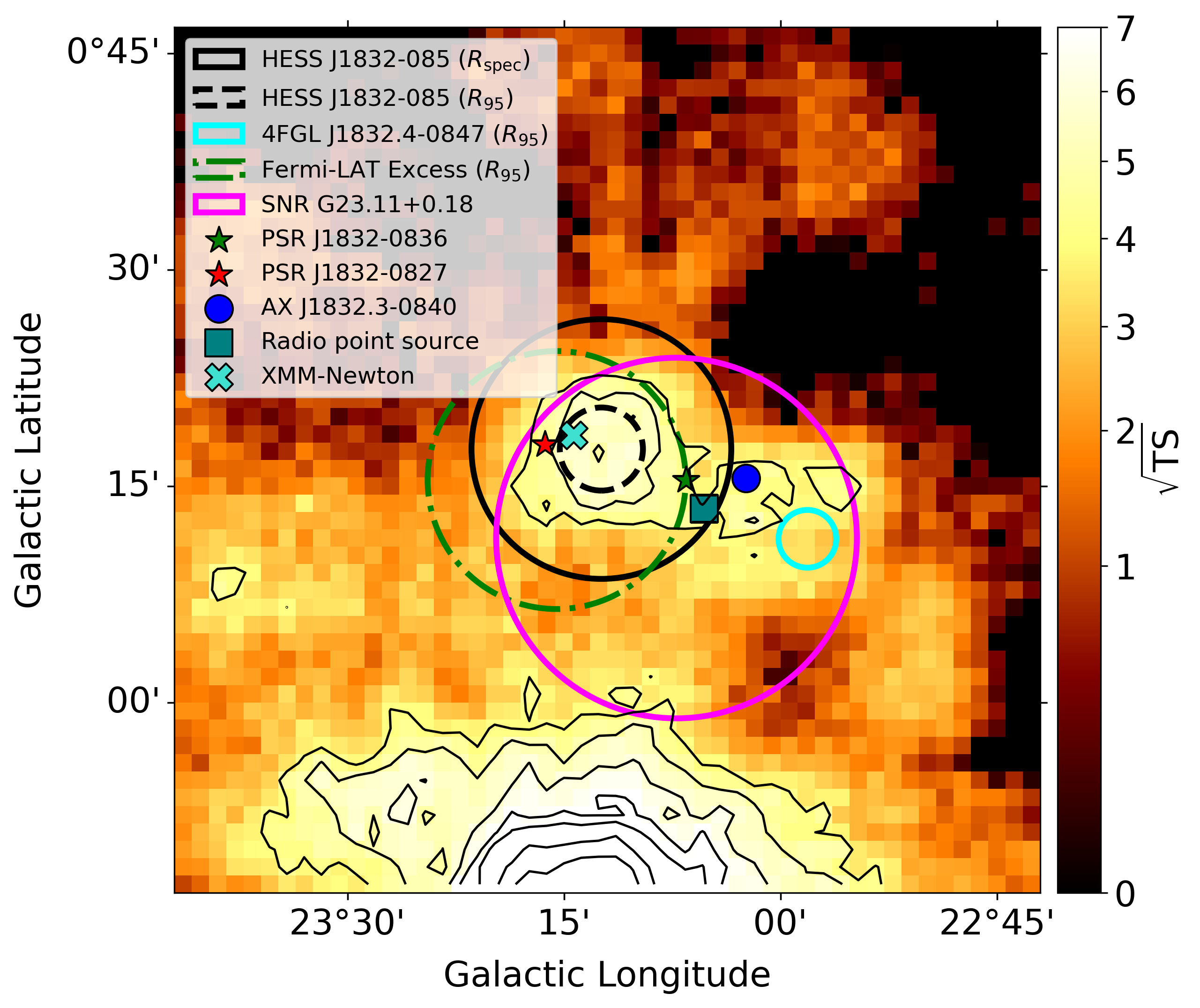}
      \caption{{\textit{Fermi}-LAT and H.E.S.S. maps of the HESS J1832-085 region.
\textit{Left panel:} \textit{Fermi}-LAT TS map in the 1-300 GeV band, computed after relocalizing 4FGL J1832.4-0847 and without including the GeV excess as a model component.
The lime-green contours correspond to TS = 7, 9, and 12, from outside to inside.
\textit{Right panel:} H.E.S.S. significance map of the region centered at HESS J1832-085.
The color bars indicate $\mathrm{TS}$ and $\sqrt{\mathrm{TS}}$ in the left and right panels, respectively.
In both panels, the black contours show the H.E.S.S. significance levels, obtained from the publicly available H.E.S.S. Galactic Plane Survey data \citep{hess18}, and trace the emission from HESS J1832-085 as well as the neighboring source HESS J1834-087, associated with the SNR W41 (G23.3-0.3), visible toward the bottom of the panels.
The solid and dashed black circles indicate the spectral extraction radius and 95\% positional uncertainty of HESS J1832-085, respectively.
The cyan circles show the 95\% statistical localization regions of 4FGL J1832.4-0847, obtained without explicitly modeling the GeV excess in the left panel and with the excess modeled as an additional point source in the right panel.
The green dash-dotted circle in the right panel represents the 95\% statistical localization region of the modeled GeV excess, centered on its best-fit position.
The magenta circle marks SNR G23.11+0.18.
PSR J1832-0827 and PSR J1832-0836 are indicated by red and green stars, respectively, while AX J1832.3-0840 is shown as a blue circle, the radio point source as a teal square, and the proposed \textit{XMM-Newton} X-ray counterpart as a turquoise cross.}}
    \label{fig: ts_change}
\end{figure*}

We analyzed $\sim$17.3 yrs of \textit{Fermi}-LAT P8R3 \texttt{SOURCE} class data (\texttt{evclass}=128, \texttt{evtype}=3; \citealt{atwood13, bruel18}) collected between 2008 August 4 (Mission Elapsed Time (MET) 239557417) and 2025 December 13 (MET 787295981) in the 1-300 GeV energy range. 
For the \textit{Fermi}-LAT analysis, we adopted a lower energy threshold of 1 GeV to reduce contamination from the bright diffuse background and to benefit from the improved angular resolution at higher energies \citep{devesa25,desarkar26}.
The LAT PSF is strongly energy dependent, with its 68$\%$ containment radius decreasing from several degrees at the lowest energies to below $0.1^{\circ}$ above 10 GeV \citep{abdollahi20}.
The analysis was performed with \texttt{Fermipy} v1.4.0 \citep{wood17}, using the \texttt{P8R3\_SOURCE\_V3} instrument response functions. 
Events with zenith angles $>90^{\circ}$ were excluded {to minimize Earth-limb contamination}, and the Galactic and isotropic diffuse backgrounds were modeled with \texttt{gll\_iem\_v07.fits} and \texttt{iso\_P8R3\_SOURCE\_V3\_v1.txt}, respectively. 
All sources from the 4FGL-DR4 catalog \citep{abdollahi20, ballet23} were included in the initial model.
{We performed a binned likelihood analysis over a $10^{\circ}\times10^{\circ}$ region of interest (ROI) {defined in Galactic coordinates and} centered on the source, with $0.1^{\circ}$ spatial binning and 8 energy bins per decade. Catalog sources within a larger $15^{\circ}\times15^{\circ}$ region centered on the same position were included in the model to account for contributions from sources outside the fitted ROI.}

Given the crowded nature of the region and nearly three additional years of \textit{Fermi}-LAT observations beyond those included in the 4FGL-DR4 catalog, we adopted the following procedure for the likelihood analysis. 
The baseline model of the ROI was optimized via \texttt{fermipy.optimize} function.
The normalizations of all catalog sources within $5^{\circ}$ of the centroid of HESS J1832-085 were left free, while all spectral parameters were freed for sources located within $2.5^{\circ}$ or having a test statistic (TS) $>500$.
All remaining source parameters were held fixed.
All parameters associated with the Galactic and isotropic diffuse components were also left free during the fit.
A search for unmodeled excesses was performed using \texttt{Fermipy}'s source-finding algorithm with the required condition of TS $>16$ and a minimum separation $0.2^{\circ}$, followed by a global refit, repeating the above procedure to stabilize the fit. 
The final model was used to derive the subsequent analysis products.

The closest 4FGL catalog source to the H.E.S.S. centroid is 4FGL J1832.4-0847, {located, after relocalization, at $(l,b)=(22.9931^{\circ} \pm 0.0133^{\circ}, 0.1927^{\circ} \pm 0.0122^{\circ})$, with an angular separation of $0.238^{\circ}$ and a detection significance of $\sim17.42\sigma$.}
No other previously undetected source was found in the region via \texttt{Fermipy}'s source-finding algorithm.
Given the large separation between this 4FGL source and the H.E.S.S. centroid, we refrain from considering this as the GeV counterpart of HESS J1832-085.
However, upon inspecting the {\it Fermi}-LAT TS map of the source region, we find a point-like GeV excess in close proximity to the X-ray source and the H.E.S.S. centroid. 
A point-like excess at a similar position was previously reported by \citet{ergin21}. 
That excess was discussed in connection with the radio SNR G23.11+0.18; however, given its {spatial offset} and point-like morphology, the GeV emission does not provide compelling morphological evidence for an association with the extended SNR.
Instead, the excess is particularly interesting in the context of a gamma-ray binary interpretation because of its spatial proximity to both the H.E.S.S. source and the proposed X-ray counterpart.
We therefore try to assess whether this residual feature could provide a more plausible GeV counterpart than the nearby 4FGL source.

The aforementioned excess found in the TS map is not recovered by the automated \texttt{Fermipy} source-finding procedure described above.
{We therefore modeled this residual feature manually as an additional point source, initialized at the position reported by \cite{ergin21}, i.e., $l = 23.2445^{\circ}$ and $b = 0.2146^{\circ}$, with a power law photon index of 2.
We then repeated the analysis following the same procedure described above.}
{The positions of both the manually added point source and 4FGL J1832.4-0847 were allowed to vary during the localization procedure.}
{Including this point source in the likelihood model yields $\mathrm{TS}=2(\ln\mathcal{L}_1-\ln\mathcal{L}_0) \approx22.4$, where $\mathcal{L}_1$ and $\mathcal{L}_0$ are the maximum likelihoods of the models with and without the excess, respectively.}
The TS map {(shown in the left panel of Fig. \ref{fig: ts_change})} was computed from the separately fitted model without an explicit component for the excess, fitting only the normalization of a test point source at each trial position, with its photon index fixed to $\Gamma=2$ and the other model parameters held fixed during the scan.
The reported excess $\mathrm{TS}\approx22.4$ instead comes from the likelihood fit of the localized source with its normalization and photon index free, with the free background parameters reoptimized under the source-absent hypothesis.
These different likelihood comparisons need not yield the same TS.
The map peak lies below the adopted source-finding threshold of $\mathrm{TS}=16$, consistent with the non-recovery of the excess by the automated search.
Through this analysis, we obtain a best-fit position of the GeV excess at $(l,b)=(23.2584^{\circ}\pm0.0682^{\circ},0.2570^{\circ}\pm0.0576^{\circ})$, with statistical localization containment radii of $R^{\rm pos}_{68}=0.0923^{\circ}$ and $R^{\rm pos}_{95}=0.1490^{\circ}$.
This best-fit position lies approximately $0.059^{\circ}$ from the H.E.S.S. centroid and $0.056^{\circ}$ from the proposed X-ray counterpart, placing both within the 95\% localization region of the excess.
With the excess included in the model, the relocalized 4FGL J1832.4-0847 lies $\sim0.261^{\circ}$ from the H.E.S.S. centroid, still disfavoring its association with HESS J1832-085 as the GeV counterpart.

During the broadband likelihood fit, assuming a power law spectrum over 1-300 GeV, we obtain a photon index of $\Gamma=2.53\pm0.22$ and an integrated energy flux of $(5.61\pm1.21)\times10^{-12}\ {\rm erg\,cm^{-2}\,s^{-1}}$ for the excess.
{The reference energy was fixed at $E_{\rm b}=1$ GeV, and the quoted uncertainties are statistical $1\sigma$ errors.}
{The best-fit power law spectrum is represented by its statistical $1\sigma$ uncertainty band in the broadband SED presented in the next section {and in Fig. \ref{fig: SED}}.}

{Given its close spatial proximity to the H.E.S.S. centroid and to the proposed X-ray counterpart, we regard the excess as a tentative GeV counterpart candidate.}
{Its moderate statistical significance prevents a robust confirmation and leaves open the possibility that the feature is related to background mismodeling or source confusion, although GeV faintness is not uncommon among gamma-ray binaries \citep{dubus17}.}
Additional \textit{Fermi}-LAT exposure, improved localization and spectral constraints, and, if an orbital period is identified, a phase-resolved analysis will be important for testing its candidacy.

\subsection{Near-Infrared counterpart candidate}\label{sec:nir}

\begin{figure}
     \includegraphics[width=\columnwidth]{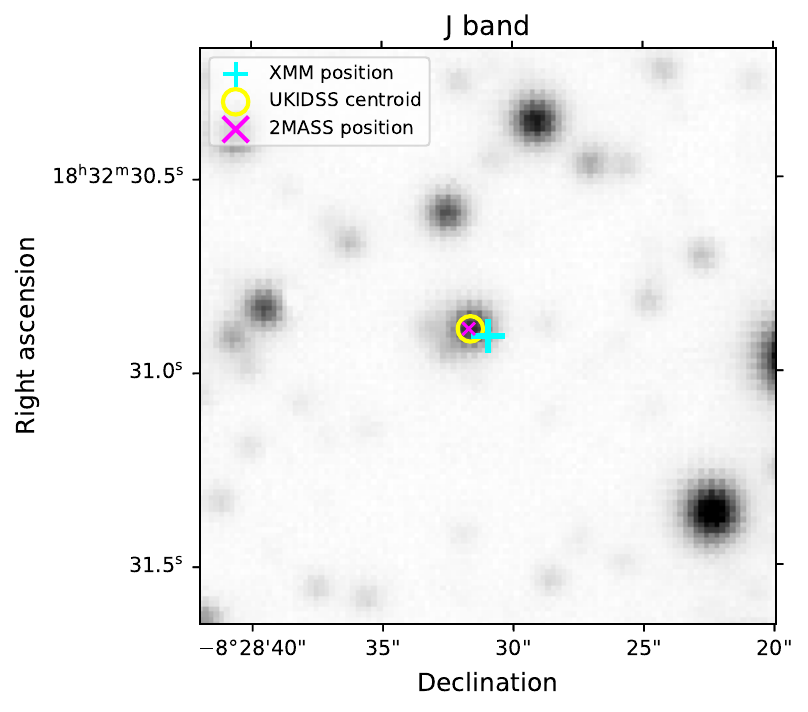}
      \caption{UKIDSS Galactic plane survey $J$-band {image} centered on the {\it XMM-Newton} source position.
The cyan plus marks the X-ray counterpart coordinate, $(l,b)=(23.2390621^{\circ},0.3090542^{\circ})$, and the yellow circle marks the centroid of the nearby NIR source.
The angular separation between the two positions is approximately {$0.73^{\prime\prime}$}.
{The magenta cross marks the catalog position of 2MASS J18323088-0828317, located $0.79^{\prime\prime}$ from the X-ray position and $0.063^{\prime\prime}$ from the UKIDSS centroid.
}}
    \label{fig: nir}
\end{figure}

We also inspected archival $J$-, $H$-, and $K$-band images from the UKIRT Infrared Deep Sky Survey (UKIDSS) \footnote{\url{http://wsa.roe.ac.uk/index.html}} \citep{lukas08} at the position of the X-ray counterpart detected by \textit{XMM-Newton}. 
The UKIDSS project is defined in \cite{lawrence07}. UKIDSS uses the UKIRT Wide Field Camera (WFCAM; \citealt{casali07}). The photometric system is described in \cite{hewett06}, and the calibration is described in \cite{hodgkin09}. The pipeline processing and science archive are described in \cite{irwin04} and \cite{hambly08}.
An unresolved NIR source is visible in all three bands, with its {$J$-band} image centroid located approximately {$0.73^{\prime\prime}$} from the X-ray counterpart position.
The source is also recovered in independent $K$-band observations obtained in 2007 and 2011, supporting its identification as a persistent astronomical source rather than an image artifact.
{We show the unsaturated UKIDSS $J$-band image in Fig. \ref{fig: nir}, together with the X-ray counterpart position, the UKIDSS centroid, and the position of the corresponding Two Micron All Sky Survey \footnote{\url{https://irsa.ipac.caltech.edu/applications/2MASS/IM/}} (2MASS; \citealt{2MASS06}) source discussed below.}
However, the field is crowded, and no formal chance-coincidence analysis or improved X-ray astrometric correction has been performed.
Moreover, the source core in the available UKIDSS $H$- and $K$-band cutouts is affected by saturation, preventing a reliable determination of NIR colors {from these images}.
We therefore identify this object only as a candidate NIR counterpart and do not attempt a stellar classification from the present data.

{For a quantitative photometric check, we use the 2MASS Point Source Catalog measurements \footnote{\url{https://irsa.ipac.caltech.edu/applications/Gator/}} of 2MASS J18323088-0828317.
Its catalog position lies $0.79^{\prime\prime}$ from the X-ray position and $0.063^{\prime\prime}$ from the UKIDSS $J$-band centroid, supporting the identification of the same NIR object in both surveys.
The catalog magnitudes are $J=14.389\pm0.058$, $H=12.110\pm0.040$, and $K_s=10.973\pm0.027$ in the Vega system.
We use the $J-K_s$ color to estimate the extinction and the $J$-band magnitude to estimate the photospheric radius, with the $H$-band measurement providing a consistency check, while retaining the UKIDSS $J$-band image for the spatial comparison.
}

{To assess the brightness and reddening expected under a hot companion interpretation, we approximate the photosphere as a blackbody with $T=3.5 \times 10^4$ K at the adopted distance of $d=4.6$ kpc.
We use the extinction law of \cite{cardelli89} with $R_V=3.1$ and evaluate the model at the nominal 2MASS wavelengths using the zero-magnitude fluxes of \cite{cohen03}.
The observed color is $(J-K_s)=3.416$, while the assumed blackbody has an intrinsic color of $(J-K_s)_0\approx-0.186$, giving a color excess of $E(J-K_s)\approx3.602$ mag.
Defining $k_\lambda=A_\lambda/A_V$, the adopted extinction law gives $k_J=0.2876$ and $k_{K_s}=0.1170$, from which we obtain $A_V=E(J-K_s)/(k_J-k_{K_s})\approx21.1 \, {\rm mag} \, \text{and} \, A_J=k_J A_V\approx6.07 \, {\rm mag}.$
The corresponding extinction-corrected absolute magnitude is $M_J=J-A_J-5\log_{10}\left(d/10\,{\rm pc}\right)\approx-5.00$.
For comparison, a $7\,R_\odot$ photosphere with the same temperature, distance, and extinction would yield $J\approx15.8$.
Using the scaling of photospheric flux with $R^2$, the observed $J$-band brightness therefore implies $R\approx13.7\,R_\odot$ at the assumed temperature and distance.
These parameters predict $H\approx12.20$, differing from the observed value by approximately $0.09$ mag.
These estimates provide an order-of-magnitude consistency check rather than a measurement of the stellar parameters, since the temperature and distance are assumed and the monochromatic blackbody approximation neglects stellar atmosphere and bandpass effects, possible blending, and non-photospheric emission.
}

The large column density inferred from the X-ray spectral analysis {is consistent with} strong line-of-sight extinction.
{Using the empirical relation $N_{\rm H}/A_V=2.87\times10^{21}\ {\rm cm^{-2}\,mag^{-1}}$ from \cite{foight16}, the NIR-inferred extinction corresponds to a dust-equivalent column of $N_{\rm H,dust}\approx6.1\times10^{22}\ {\rm cm^{-2}}$.
This model-dependent estimate is broadly consistent with the column density $N_{\rm H}=(8.0^{+2.5}_{-2.0})\times10^{22}\ {\rm cm^{-2}}$ obtained from the \textit{XMM-Newton} spectrum, considering the quoted X-ray uncertainties and the approximate nature of the NIR extinction estimate.
}
Optical and ultraviolet emission from a putative companion would therefore be strongly absorbed {under the inferred reddening}, whereas the extinction becomes progressively weaker toward the NIR, particularly in the $K$ band.

{For an indicative optical extrapolation, we adopt a representative hot companion intrinsic color $(I_C-J)_0\approx-0.4$ \citep{pecaut13} and approximate $A_{I_C}/A_V\approx0.60$ using the extinction law of \cite{cardelli89} at
$\lambda\approx0.80\,\mu{\rm m}$ for $R_V=3.1$. These assumptions give $I_C=J+(I_C-J)_0+A_{I_C}-A_J\approx20.6.$
This magnitude is a model prediction, not an optical measurement or an observational upper limit, and testing it requires an optical measurement or a reliable local detection limit.
}

A luminous early-type companion, potentially an O- or Be-type star, would be qualitatively compatible with the proposed binary interpretation.
Nevertheless, the current {imaging and broadband photometry} cannot determine the spectral type, luminosity class, or presence of a circumstellar component, nor can they exclude an unrelated foreground or background star.
{Higher-resolution, unsaturated} multi-band photometry and NIR spectroscopy will be required to establish the nature of the candidate counterpart.

\section{Discussion}\label{discussion}

\begin{figure*}
    \centering
    \includegraphics[width=\textwidth]{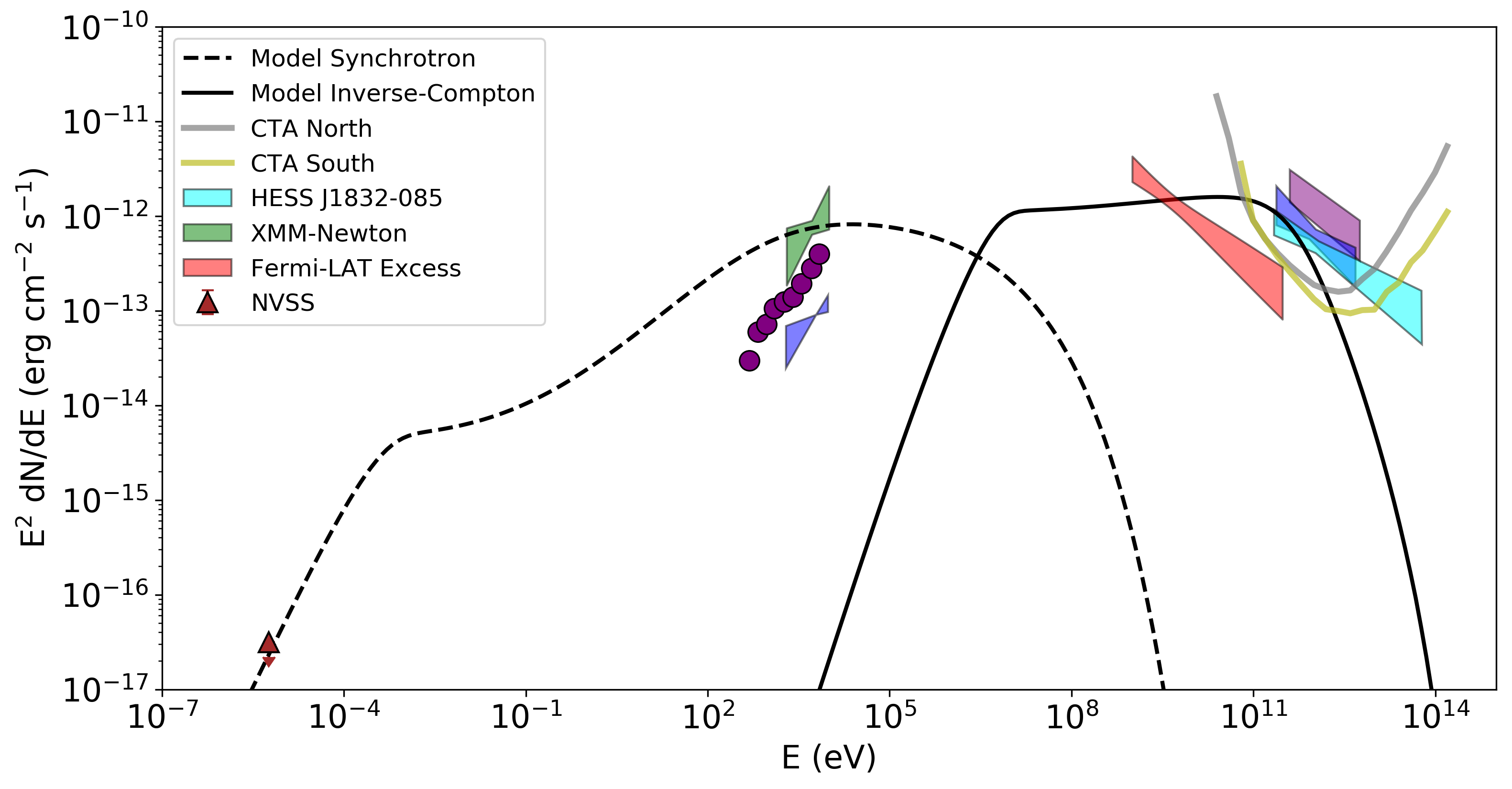}
    \caption{Broadband SED of HESS J1832-085, along with the model synchrotron and IC emission. 
    The X-ray and TeV data for HESS J0632+057 \citep{aharonian07, hinton_09} and HESS J1832-093 \citep{hess15a, Eger_16, Mart_2020} are also shown in all purple and all blue, respectively, for comparison. 
    }
    \label{fig: SED}
\end{figure*}

\begin{figure}
     \includegraphics[width=\columnwidth]{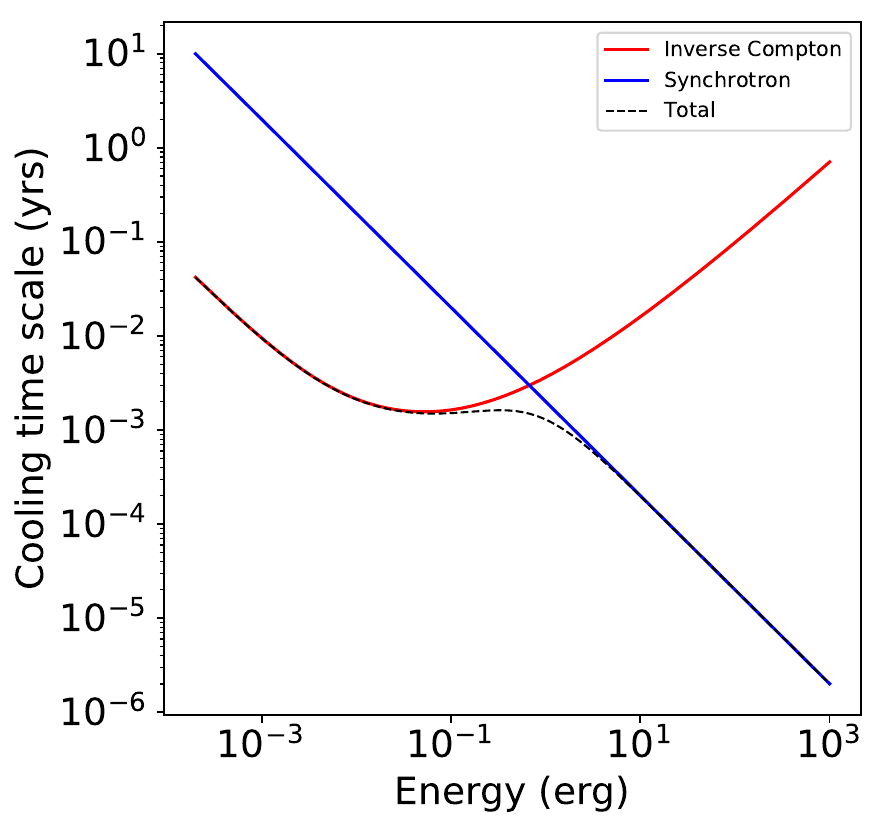}
      \caption{{{Energy-dependent radiative cooling of the modeled electron population. The {curves show} the synchrotron, IC, and total cooling times {as functions of electron energy}.}}}
    \label{fig: loss}
\end{figure}

In this section, we discuss whether the spectral properties of HESS J1832-085 support its classification as a candidate gamma-ray binary.
The X-ray spectrum, derived from the \textit{XMM-Newton} observation, is hard, strongly absorbed, and non-thermal.
The hard spectral slope is compatible with those commonly measured in established gamma-ray binaries (see, e.g., \citealp{hinton_09, Eger_16}).
The system is also strongly absorbed, and the column density obtained from \textit{XMM-Newton} analysis is very much consistent with that obtained for the X-ray counterpart of HESS J1832-093 \citep{Eger_16, Mart_2020}.

A defining feature of a confirmed gamma-ray binary would be the presence of orbital modulation, particularly in the X-ray band, whereas the same in the GeV band is difficult to attain \citep{Tam_20, desarkar22a}.
We did not perform a GeV orbital variability analysis because the limited significance and photon statistics of the excess would make a blind periodicity search poorly constrained.
The available GeV data therefore do not provide an orbital periodicity test of the binary interpretation.
The lack of a GeV orbital periodicity constraint does not, by itself, argue against a binary interpretation, since the modulation amplitude can be weak or difficult to recover depending on the orbital inclination, eccentricity, viewing geometry, and available photon statistics.
Nevertheless, the long-term light curve produced from available X-ray data (see Fig. \ref{fig: xray_variab}) hints towards a flux variability, which is qualitatively similar to the behavior observed in other gamma-ray binary systems (see Fig. 4 of \citealp{Eger_16}).
Long-term monitoring of the source is therefore essential; especially regular X-ray observations with \textit{Swift} or comparable facilities would provide the probe for orbital periodicity, given the typically stronger modulation observed in X-rays.

{In the following, we consider the possibility that the X-ray source studied here is the counterpart of HESS J1832-085.
Other X-ray sources are present within the H.E.S.S. positional uncertainty region, and this association therefore remains tentative.}
The broadband SED of HESS J1832-085 is shown in Fig. \ref{fig: SED}.
From the plot, it can be seen that the TeV and X-ray spectra of the respective counterparts show remarkable similarity with those of the established gamma-ray binaries.
Specifically, a hard, strongly absorbed spectrum is observed in X-ray, whereas a softer spectrum ($\Gamma = 2.38 \pm 0.14$) is observed in the TeV range \citep{hess18}, typical of gamma-ray binaries.
We further try to explain the broadband SED by solving the one-zone diffusion-loss equation using \texttt{GAMERA} \citep{hahn15, hahn22}, considering a single leptonic population, assumed to be accelerated {in the wind collision region}. 
{At radio frequencies, no compact counterpart is detected at the \textit{XMM-Newton} position. 
We therefore use the NVSS constraint \citep{condon98} as an upper limit on any compact radio emission associated with the candidate gamma-ray binary. 
This constraint should be distinguished from the radio emission associated with the extended SNR G23.11+0.18 reported by \cite{maxted19}. 
The latter traces diffuse emission from the larger-scale remnant environment, measured over a substantially larger angular region, and is therefore not directly comparable to the compact source limit adopted here. 
In particular, using the integrated SNR radio flux in the broadband model would incorrectly attribute large-scale shell emission to the putative compact binary counterpart. 
For this reason, we retain the NVSS upper limit as the appropriate radio constraint for the compact source; this limit constrains the minimum energy of the lepton population in the broadband modeling.}

The simple modeling discussed here is illustrative, and follows similar arguments to \cite{hinton_09}.
Accelerated electrons are assumed to be injected with a power law spectrum with an exponential cutoff, $dN/dE \propto E^{-\alpha} \exp (-E/E_c)$ with $\alpha = 2$ and $E_c = 50$ TeV, with a minimum energy of $E_{\rm min} \approx 0.1$ GeV.
We further {adopt an illustrative} strong and {blue radiation field from the putative companion star} with temperature $T=3.5\times10^4$ K ($k_{\rm B}T\sim3$ eV) and stellar {photon} energy density
$U\approx0.1\ {\rm erg\,cm^{-3}}$ {at the emission region}.
{The temperature is representative of a hot early-type companion and is also assumed in the NIR consistency check in Subsect. \ref{sec:nir}, rather than independently determined by that analysis.
The energy density depends on the stellar luminosity and the distance between the star and the emission region, neither of which is independently established for this system.
These parameters therefore specify an illustrative radiation field without determining the companion's luminosity class or the system geometry.}
The IC upscattering occurs in the deep Klein-Nishina (KN) regime in the case of TeV gamma-ray emission.
{Assuming that the X-ray and gamma-ray emission arise predominantly from synchrotron and IC radiation, respectively, from the same underlying electron population, the relative emission levels imply a magnetic field of $B \le \sqrt{8\pi f_{\rm KN} U (F_X/F_{\gamma})} \ \mathrm{G} \sim 100$ mG, where $f_{\rm KN}$ accounts for the suppression of IC scattering in the KN regime.
For $k_{\rm B}T\sim3$ eV and $E=1$ TeV, $f_{\rm KN}\approx [1+(4E\times2.8k_{\rm B}T)/(m_ec^2)^2]^{-1.5} \sim10^{-3}$ \citep{moderski05}.}
The {adopted model parameters are comparable to those} discussed in \cite{hinton_09} for HESS J0632+057, and {those} subsequently used for HESS J1832-093 \citep{Eger_16, Mart_2020}.
Assuming an association with SNR G23.11+0.18, which likely resulted from the explosion that created the compact object in the binary, we consider the same age of $t_{\rm age} \approx10^4$ yrs and distance of $d \approx4.6$ kpc as that of the SNR \citep{maxted19}.
Finally, to explain the SED, the required luminosity in electrons is $L_e \approx 6.5 \times 10^{34} \ \rm erg \ s^{-1}$, which is comparable to that obtained by \cite{hinton_09} for HESS J0632+057, and is compatible with the available kinetic power of the Be-star wind \citep{waters87}. 
{As shown in Fig. \ref{fig: SED}, the resulting model broadly reproduces the X-ray and TeV emission.
However, the model fails to reproduce the nominal power law shape inferred for the tentative GeV excess.
Note that the photon index of the tentative GeV excess is consistent within the uncertainties with the phase-averaged value $\Gamma=2.40\pm0.06_{\rm stat}\pm0.06_{\rm sys}$ reported for the proposed GeV counterpart of HESS J0632+057 by \cite{Li_17}.
In that system, the GeV power law also does not extrapolate directly to the TeV spectrum.
This comparison illustrates that different GeV and TeV spectral shapes can occur in gamma-ray binaries, but does not establish the compatibility of our particular model with the \textit{Fermi}-LAT data.
}

{Fig. \ref{fig: loss} shows that the modeled electron population is strongly loss-dominated. 
The total radiative cooling time remains much shorter than the assumed system age of $\sim10^{4}$ yr throughout essentially the entire injected energy range. 
The present-day electron distribution therefore does not represent particles accumulated over the lifetime of the system, but instead approaches a loss-controlled state determined by the balance between continuous injection and radiative cooling. 
In this regime, the adopted age mainly ensures that the cooled particle distribution has been established, while the HE population is governed by the radiative loss rates.

{The energy dependence of the radiative losses} directly shapes the evolved electron spectrum.
Where the total loss rate is large, electrons are removed rapidly from the corresponding energy range, and the particle population is depleted relative to the injected spectrum. 
At higher energies, when IC cooling becomes less efficient in the KN regime, electrons retain their energy for longer than would be expected from an extrapolation of the lower-energy cooling behavior.
This produces a change in the curvature of the evolved electron distribution rather than a single uniformly cooled power law spectrum. 
At the highest energies, the increasing importance of synchrotron losses again causes strong depletion of the electron population and reinforces the decline toward the imposed cutoff. 
Thus, the modeled X-ray and TeV emission is determined not only by the injected electron power law index and cutoff, but also by the full energy-dependent competition between IC and synchrotron losses included in the calculation.}

\cite{maxted19} have previously proposed that TeV emission from HESS J1832-085 could arise from the spatially coincident SNR G23.11+0.18 interacting with a nearby molecular cloud (MC).
This is not unlike HESS J1832-093, where SNR G22.7-0.2 was initially classified as the responsible HE emitter \citep{acero16}.
\cite{hess15a} argued that the positional offset between SNR G22.7-0.2 and HESS J1832-093, and the presence of infrared and X-ray sources spatially compatible with the gamma-ray emission, strongly point against the SNR origin.
A similar case can be argued for HESS J1832-085 as well.
The moderate positional offset between the H.E.S.S. source and the SNR, the point-like TeV morphology, the presence of a hard non-thermal X-ray source close to the TeV centroid, indications of its long-term X-ray variability, {along with a candidate NIR counterpart} collectively argue against the TeV emission being solely dominated by hadronic processes in an SNR+MC system, and instead favor a compact binary origin. 
The broadband SED modeling further supports this interpretation, as the observed {X-ray and TeV emissions} are more naturally explained within a binary scenario, {although the GeV emission remains unexplained by the adopted model.}
{Note that this phenomenology is also very typical of gamma-ray binaries, and has been explained using a separate population of leptons accelerated by Coriolis turnover, as a result of orbital motion \citep{zabalza13}.}
A more complex emission scenario cannot be excluded, such as a two-zone structure responsible for the X-ray and TeV emission, or a compact binary component superimposed on a more extended contribution arising from an SNR+MC interaction. 
A detailed investigation of such hybrid configurations is beyond the scope of this work and is deferred to future studies.
We further note that the nearby pulsars, the radio point source, and the IP candidate in the field are unlikely to be physically related to the TeV source, either due to positional offsets, incompatible energetics, or spectral properties that do not match the observed HE emission.

Given the limited available data, the archival X-ray observations may have captured the source during a relatively high flux state, highlighting the need for simultaneous MWL observations to probe X-ray$-$gamma-ray correlations and constrain variability timescales. 
Overall, {assuming that the proposed X-ray association is correct,} the current MWL evidence favors HESS J1832-085 as a promising gamma-ray binary candidate, with its compact TeV morphology, {candidate X-ray counterpart with indications of variability}, {a nearby candidate NIR counterpart that could represent the stellar companion,} and broadband spectral properties pointing toward a binary origin.
Definitive confirmation will require long-term phase-resolved X-ray monitoring to establish orbital modulation, together with deep infrared observations to identify the stellar counterpart and better constrain the non-thermal emission environment.

\begin{acknowledgements}
We thank the anonymous reviewer for helpful suggestions and constructive criticism, which vastly improved the quality of this
work.
      ADS thanks Ömer Faruk \c{C}oban and Guillem Martí-Devesa for helpful discussion regarding \textit{Fermi}-LAT data analysis.
      ADS is supported by the grant Juan de la Cierva JDC2023-052168-I, funded by MCIU/AEI/10.13039/501100011033 and by the ESF+.
      This work has been supported by the grant PID2024-155316NB-I00 funded by MICIU /AEI /10.13039/501100011033 / FEDER, UE, and CSIC PIE 202350E189. This work is also partly supported by the Spanish program Unidad de Excelencia María de Maeztu CEX2020-001058-M, financed by MCIN/AEI/10.13039/501100011033, and by the MaX-CSIC Excellence Award MaX4-SOMMA-ICE, and also supported by MCIN with funding from European Union NextGeneration EU (PRTR-C17.I1).
\end{acknowledgements}

%
   \bibliographystyle{aa} 
   \bibliography{aa.bib} 
%

\end{document}